\documentclass[11pt]{article}
\usepackage{latexsym}
\usepackage{epsfig}

\def\bg#1{\mbox{\boldmath$#1$}}

\newcommand{\anabla}{{\overrightarrow{\nabla}}\!\!\!\!\!\!{\overleftarrow{\nabla}}}

\newcommand{\del}{\partial}
\newcommand{\beq}{\begin{eqnarray}}
\newcommand{\eeq}{\end{eqnarray}}
\newcommand{\be}{\begin{eqnarray*}}
\newcommand{\ee}{\end{eqnarray*}}

\newcommand{\br}{{\bf r}}

\newcommand{\e}{\epsilon}

\begin{document}

\centerline{\Large\bf {Proton-Proton Scattering Lengths from Effective Field Theory}}
\vskip 10mm
\centerline{Xinwei Kong$^1$ and Finn Ravndal\footnote{On leave of absence from Institute 
            of Physics, University of Oslo, N-0316 Oslo, Norway}} 
\medskip
\centerline{\it Department of Physics and Institute of Nuclear Theory,}
\centerline{\it University of Washington, Seattle, WA 98195, U.S.A}

\bigskip
\vskip 5mm
{\bf Abstract:} {\small Using a recently developed effective field theory for
the interactions of nucleons at non-relativistic energies, we calculate 
the Coulomb corrections to proton-proton scattering. Including the dimension-eight 
derivative interaction in PDS regularization scheme, we obtain a modified Jackson-Blatt 
relation for the scattering lengths which is found to be phenomenologically satisfactory.
The effective range is not modified by Coulomb effects to this order in the calculation.}


\bigskip
During the last year Kaplan, Savage and Wise have proposed an effective field theory for the 
interactions of nucleons at low energies\cite{KSW_1}. These are characterized by scattering 
lengths which are much larger than the natural hadronic scale which in these systems
is set by the pion mass. This made it initially difficult to construct consistent power
counting rules which are necessary in any effective theory calculation to ensure that all 
contributions to a given order are included. By the introduction of a new regularization
scheme called Power Divergence Subtraction (PDS) which is an extension of the more common
MS scheme, these problems were solved. A very similar scheme was proposed at the same
time by Gegelia\cite{Gegelia}. Very recently, Mehen and Stewart\cite{Tom} have made
this off-shell scheme (OS) more well-defined and shown that it is in fact equivalent to the
PDS scheme. 

Within this framework several applications to problems involving the deuteron have now been
made\cite{KSW_2}\cite{gamma}. The first attempts to describe three-nucleon systems with two
neutrons have also been initiated\cite{Paulo}. In this way the properties of the triton 
nucleus can be investigated. But also systems like $^3He$ with two protons are of obvious 
interest to consider in this new approach. For this to succeed, one must know how to 
include the repulsive Coulomb force between the two protons. As a first step in this 
direction we will here consider proton-proton scattering at very low energies.

The effective Lagrangian for non-relativistic protons with mass $M$ in the spin-singlet 
channel is
\beq
    {\cal L}_0 = p^\dagger\left(i\del_0 + {{\bg{\nabla}}^2\over 2M}\right)p 
             - {C_0\over 2}(p\sigma_2p)(p\sigma_2 p)^\dagger                  \label{Leff}
\eeq
when we only include the lowest order contact interaction parameterized by the coupling
constant $C_0$. It corresponds to the singular potential $C_0\delta(\br)$ which will
affect interactions only in the $S$-wave. In addition, we must also include the static 
Coulomb repulsion between the
protons. It has the effective strength $\eta = \alpha M/2p$ where $p$ is the CM momentum of
protons, and thus also becomes strong at very low energies. This can be done using the
well-established formalism based upon the exact solutions of the Schr\"odinger equation
in the Coulomb potential\cite{Coulomb}. Denoting these  by $\psi({\br})$, the probability to 
find the protons at the same point in the initial or final state is reduced and 
given by $|\psi(0)|^2 =  C_\eta^2$ where
\beq
     C_\eta^2 = {2\pi\eta\over e^{2\pi\eta} - 1}                   \label{Sommer}
\eeq
is the Sommerfeld factor\cite{Coulomb}. In a partial wave expansion of the full scattering 
amplitude\cite{GW}, the total phaseshift 
is $\sigma_\ell + \delta_\ell^C$ where $\sigma_\ell$ is the pure Coulombic shift and 
$\delta_\ell^C$ is the shift due to the strong interactions. If we denote this latter one 
in the $S$-wave by $\delta_{pp}^C$, it is related to the corresponding strong amplitude
$T_{SC}(p)$ by the standard partial wave expression
\beq
    p\,(\cot\delta_{pp}^C - i) = - {4\pi\over M} {e^{2i\sigma_0}\over T_{SC}(p)} \label{Ccot}
\eeq
The superscript $C$ is a reminder to the fact that this is not the strong phaseshift in the
absence of Coulomb effects. Instead it is the strong phaseshift modified by the Coulomb 
potential. It is only the Coulomb repulsion between the protons in the initial and final
states that have been removed at this stage.

As has been known for a long time, $\cot\delta_{pp}^C$ in (\ref{Ccot}) does
not have a regular effective range expansion\cite{Bethe}. One finds instead
\beq
    p\left[C_\eta^2\left(\cot\delta_{pp}^C - i\right) + 2\eta H(\eta)\right]
    = - {1\over a_{pp}^C} + {1\over 2}r_0\,p^2 + \ldots                      \label{CERE}
\eeq
where $a_{pp}^C$ and $r_0$ is respectively the $S$-wave Coulomb-modified scattering 
length and effective range. They arise after removing the part of the amplitude described 
by the complex function\cite{Haeringen}
\beq
     H(\eta) = \psi(i\eta) + {1\over 2i\eta} - \ln(i\eta)                       \label{Hfun}
\eeq
which represents Coulomb effects at short distances. Its imaginary part cancels 
the unitary term $\sim i$ in (\ref{CERE}). The real
part defines the function  $h(\eta) = \mbox{Re}\psi(i\eta) - \ln\eta$ which is more 
suitable for phenomenological analysis\cite{BJ}.

By this approach one were able to extract the two scattering parameters a long time 
ago, $a_{pp}^C = -7.82\,\mbox{fm}$ and the effective range $r_0 = 2.83\,\mbox{fm}$ \cite{BJ}.
The scattering length is in magnitude much smaller than for $pn$  
and $nn$ scattering in the spin singlet channel where one finds respectively the
values $a_{pn} = -23.75\,\mbox{fm}$ and $a_{nn} = -18.8\,\mbox{fm}$. The difference
between these two latter numbers is well within the range determined by the breaking of
isospin invariance in strong interactions\cite{Ernest}. For proton-proton scattering on
the other hand it became clear that not all short-distance effects of the Coulomb potential
were taken out by the $H$-function in (\ref{CERE}). By investigating the detailed behavior
of the Schr\"odinger equation for the coupled problem of strong and Coulomb forces 
at small separations Jackson and Blatt\cite{JB} succeeded in finding an expression for the 
scattering length $a_{pp}$ due to only the strong interactions,
\beq
    {1\over a_{pp}} =  {1\over a_{pp}^C} + \alpha M
    \left[\ln {1\over\alpha Mr_0} - 0.33\right]                       \label{BJapp}
\eeq
With the above value for the effective range $r_0$ we then find $a_{pp}= -17.0\,\mbox{fm}$
which is very close to the $nn$ scattering length. The term $-0.33$ in the above formula is
found to be only weakly dependent on the exact form of the strong 
potential\cite{Heller}.

From the effective Lagrangian (\ref{Leff}) we can now calculate the proton-proton scattering
amplitude. It will consist of the sum of all chains of proton bubbles separated by the
vertex with value $C_0$.
Including the Coulomb potential corresponds to having static photon exchanges between all
proton lines in these Feynman diagrams. The Coulomb interactions in the initial and final
states just give the Sommerfeld factor (\ref{Sommer}). In perturbation theory each bubble
will be replaced by an infinite sum of bubbles with zero, one, two and so on Coulomb 
interactions as shown in Fig.1. 

\begin{figure}[htb]
 \begin{center}
  \epsfig{figure=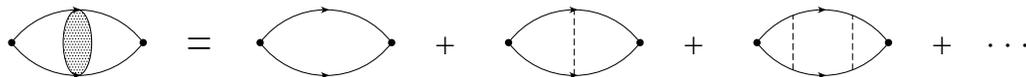,height=10mm}
 \end{center}
 \vspace{-5mm}
 \caption{\small The Coulomb-dressed bubble is a sum of bubbles with static Coulomb exchanges.}        
 \label{fig1}
\end{figure}

\begin{figure}[tb]
 \begin{center}
  \epsfig{figure=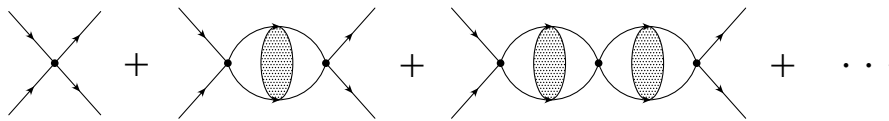,height=15mm}
 \end{center}
 \vspace{-5mm}
 \caption{\small The proton-proton scattering amplitude is a sum of chains of Coulomb-dressed 
                 bubbles.}
 \label{fig2}
\end{figure}

Since the scattering amplitude represented by the diagrams
in Fig.2 forms a geometric series, it thus takes the value
\beq
     T_{SC}(p) =  C_{\eta}^2{C_0\,e^{2i\sigma_0}\over 1 - C_0\,J_0(p)}    \label{TSC}
\eeq
Here $J_0(p)$ is the value of one bubble which equals the amplitude for a pair of 
protons with energy $E = p^2/M$ to move from zero separation to zero separation. 
This is given exactly by the the Coulomb 
propagator $G_C(E;\br' = 0,\br = 0)$. In momentum space it becomes
\beq
      J_0(p) = M\!\int\!{d^3 k\over (2\pi)^3} {2\pi\eta(k)\over e^{2\pi\eta(k)} - 1}
      {1\over p^2 - k^2 + i\e}                                            \label{Cbubble}
\eeq
When this result for the scattering amplitude is now  used in (\ref{Ccot}) and 
(\ref{CERE}), we see that both the phaseshift $\sigma_0$ and the Sommerfeld factor 
$C_0^2(\eta)$ cancel out. We are thus only left with the evaluation of (\ref{Cbubble}) which
we do using the PDS regularization scheme\cite{KSW_1} in $d = 3 - \e$ dimensions and with
a renormalization mass $\mu$. It is ultraviolet divergent which shows up as an $1/\e$ pole in 
the integral. This will be cancelled by
counterterms which renormalize the coupling $C_0$ in (\ref{TSC}) to $C_0(\mu)$. As a result,
the finite part of the dressed bubble (\ref{Cbubble}) is found to be
\beq
      J_0^{finite}(p) = {\alpha M^2\over 4\pi}\left[ \ln{\mu\sqrt{\pi}\over\alpha M}
                  + 1 - {3\over 2}C_E - H(\eta) \right] - {\mu M\over 4\pi}           \label{J0fin}
\eeq                       
where $C_E = 0.5772\cdots$ is Euler's constant. The last term here is the contribution 
from the special PDS pole in $d=2$ dimensions. We see
now that also the function $H(\eta)$  cancels out in (\ref{CERE}). There is no
contribution to the effective range $r_0$ to this order in the effective theory. Defining
the $\mu$-dependent strong scattering length
\beq
    {1\over a_{pp}(\mu)} = {4\pi\over MC_0(\mu)} + \mu                       \label{app}
\eeq
as for $pn$ and $nn$ interactions\cite{KSW_1}, we thus see from (\ref{CERE}) and (\ref{TSC}) 
that it can be expressed in terms of the measured scattering length $a_{pp}^C$ as
\beq
    {1\over a_{pp}(\mu)} =  {1\over a_{pp}^C} + \alpha M
    \left[\ln{\mu\sqrt{\pi}\over\alpha M} + 1 - {3\over 2}C_E\right]             \label{Capp}
\eeq
This relation has the exact same form as the Jackson-Blatt relation (\ref{BJapp}) where
now the renormalization length $1/\mu$ enters instead of the effective range. But 
$a_{pp}(\mu)$ is not a physical quantity in the sense that it can be measured directly and 
will thus in general depend on the renormalization point $\mu$. Coulomb effects on length 
scales $ > 1/\mu$ have been removed from it. One would expect the coupling constant 
$C_0(\mu)$ to be within a few percent of the corresponding coupling constants for $pn$ and 
$nn$ scattering. 
Our result is non-perturbative both in the strong coupling and in the fine structure 
constant which is seen to enter in the combination $\alpha\ln\alpha$. This is a
consequence of the Coulomb force becoming strong when the energy goes to zero. 

\begin{figure}[htb]
 \begin{center}
  \epsfig{figure=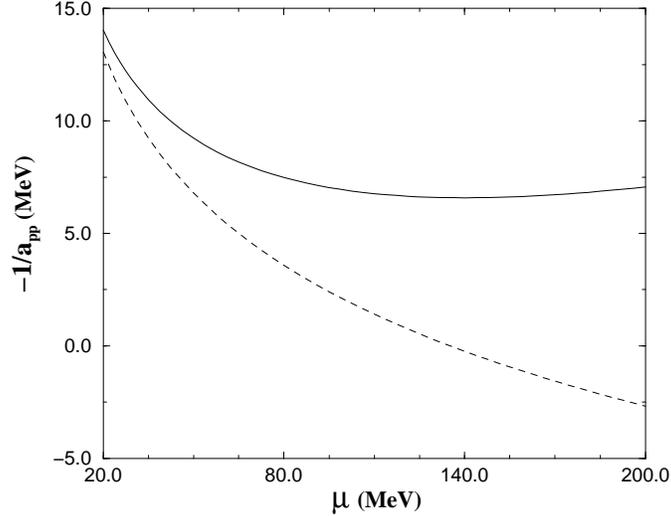,height=80mm}
 \end{center}
 \vspace{-8mm}
 \caption{\small Dependence of the inverse scattering length on the renormalization point.
                 Dashed curve gives result with only $C_0$ interaction, while the solid curve
                 also includes the $C_2$ interaction.}        
 \label{fig3}
\end{figure}

Since we are not including pions in the present effective theory, we must have the
renormalization mass in the range $|1/a_{pp}^C| < \mu \le m_\pi$. Depending on the exact
value  we see that the Coulomb correction can actually become of the same magnitude as the 
strong interaction. In fact, since the scattering length $a_{pp}^C$ is negative, it can have 
a very big effect on the hadronic scattering length $a_{pp}$. This is seen from the
right-hand side of (\ref{Capp}) which becomes zero when $\mu \simeq m_\pi$ as shown in Fig.3.
It is not a very physical result, corresponding to the hadronic scattering length 
$a_{pp}$ being infinite.

The exact numerical properties of our result is obviously dependent on the regularization
scheme we use. As a check, we have also calculated the Coulomb-dressed bubble (\ref{Cbubble})
using instead a momentum cutoff $\Lambda$ as a regulator. It will then appear in the 
logarithmic term of (\ref{Capp}) instead of $\mu$ and the numerical constants inside the 
parenthesis will be different. In fact, the new result can be obtained from (\ref{Capp}) 
by letting $\mu \approx 0.56 \Lambda$. This corresponds to a blowup of the hadronic 
scattering length for $\Lambda \approx 240\,\mbox{MeV}$.

From a physical point of view such an unstability in the result is not acceptable. Instead
we should like to see the dependence on the regulator mass to be as small as possible. The
obvious way to cure this instability is to include higher order interactions in the effective
Lagrangian (\ref{Leff}). In the $S$-wave channel there is only one such interaction,
\beq
     {\cal L}_2  = {1\over 4}C_2(p\sigma_2\anabla^2 p)(p\sigma_2 p)^\dagger + h.c.  \label{C2}
\eeq
where the operator $\anabla = 1/2(\overrightarrow{\nabla} - \overleftarrow{\nabla})$. 
Kaplan, Savage and Wise\cite{KSW_1} showed that it can be included in a perturbative way 
and gives directly the effective range $r_0$ of the proton. The resulting Feynman diagrams
can be obtained from the original diagrams by replacing one $C_0$ vertex in each diagram 
in Fig.2 by a $C_2$ vertex in every possible way. We then find the following correction
to the scattering amplitude
\beq
     \delta T_{SC}(p) = -{C_2 e^{2i\sigma_0}\over [1 - C_0 J_0(p)]^2}\psi^*(0)\nabla^2\psi(0)
         + \ldots                                                 \label{deltaT}
\eeq
when we neglect higher orders in the fine structure constant. It depends crucially on the
properties of the Coulomb wavefunction at the origin. While $|\psi(0)| = C_\eta$, the double
derivative is divergent which is easily seen from the Schr\"odinger equation. A regulated
expression can be obtained from the exact wavefunction which we transform to momentum space. 
The resulting integration is then done in the PDS scheme. In this way we find 
\beq
     \psi^*(0)\nabla^2\psi(0)|_{reg} = C_\eta^2\left(- p^2 + \alpha M\mu 
     + {1\over 2}\alpha^2 M^2\right)
\eeq
after a rather elaborate calculation. The term proportional to $\mu$ is again from a PDS 
pole. Dropping the higher order $\alpha^2$ term, we thus have the following correction to 
the inverse scattering amplitude
\beq
  \delta \left({1\over T_{SC}(p)}\right) = {e^{-2i\sigma_0} C_2(\mu)\over C_\eta^2C_0^2(\mu)}
                                        \left(p^2 - \alpha M\mu\right)      \label{dinvT}
\eeq
Comparing with the effective range expansion (\ref{CERE}) we then obtain for the effective
range of the proton simply $r_0 = 8\pi C_2/ MC_0^2$. It is therefore not modified by
the Coulomb interactions. This is consistent with experiments which show that the effective
range for $pp$ is the same as for $pn$ and $nn$ to within a few percent\cite{Ernest}.

With the coupling constant $C_2$ determined in this way, we can now
calculate its contribution to the scattering length by taking the limit of zero momentum
in (\ref{dinvT}). The PDS pole is seen to give an extra term in the previous result
(\ref{Capp}) which now is changed into
\beq
    {1\over a_{pp}(\mu)} =  {1\over a_{pp}^C} + \alpha M
    \left[\ln{\mu\sqrt{\pi}\over\alpha M} + 1 - {3\over 2}C_E - \frac12 \mu r_0\right]  \label{C2app}
\eeq
Although this new term has been calculated in perturbation theory, it comes into the result
with the same general magnitude as the first Coulomb correction and actually stabilizes the
result when considered as a function of the renormalization point $\mu$. This is shown in
Fig.3 where the corresponding curve varies little  
in the whole interval $80\,\mbox{MeV} < \mu < 200\,\mbox{Mev}$. 
The blowup of the scattering 
length for $\mu$ approaching the pion mass has now been turned into a rather smooth behaviour
in the same region. In fact, we find $-1/a(\mu = m_\pi) = 6.6\,\mbox{MeV}$. For comparison, it should be noted that $-1/a_{pn} = 8.3\,\mbox{MeV}$ 
and $-1/a_{nn} = 10.5\,\mbox{MeV}$ while (\ref{BJapp}) gives $-1/a_{pp} = 11.6\,\mbox{MeV}$. 
These differences in scattering lengths correspond to much smaller differences in the 
corresponding coupling constants $C_0$. The strong scattering length in our approach is 
scheme dependent and it will not enter explicittly in any physical results.




We would like to thank Martin Savage and all other members of the EFT group for much help and
many useful discussions. In addition, we want to thank the Department of Physics and the 
INT for generous support and hospitality. Xinwei Kong is supported by the Research Council
of Norway.

\end{document}